# Woodpile and diamond structures by optical interference holography

*Wing Yim Tam**

Department of Physics and Institute of Nano Science and Technology
Hong Kong University of Science and Technology
Clear Water Bay, Kowloon, Hong Kong, China

**Abstract**

We report the use of an optical interference holographic setup with a five-beam configuration, consisting of four side beams and one central beam from the same half space, to fabricate woodpile and diamond structures for the use as photonic bandgap materials in which electromagnetic waves are forbidden in the bandgap. By exploiting the advantage of the binarization of the interference pattern, using intensity cut-off, either linear or circular central beam can be used. More importantly, the beam configurations can be easily implemented experimentally as compared to other configurations in which the interfering beams are counter-propagating from both half spaces.

* Corresponding Author: phtam@ust.hk; Phone: 852-2358-7490; Fax: 852-2358-1652.

Photonic crystals are dielectric materials that exhibit photonic bandgaps in which electromagnetic wave propagation in the bandgaps is forbidden, analogous to electronic bandgaps in semi-conductors.[1] They hold the promises for the next generation advances in science and technology, the same as what semi-conductors had done in the past few decades. The design and fabrication of photonic crystals exhibiting complete bandgaps have been challenging, and thus have attracted wide interest.[1,2] Among the various micro-structures, the diamond and the related woodpile structures stand out with wide and complete bandgaps even with moderate dielectric contrast.[2] To fabricate the photonic crystals, methods like the self-assembly or nano-manipulation of colloidal micro-spheres,[3,4] the layer by layer micro-fabrications,[5] and recently, the holographic lithography[6-8] and the multi-photon direct laser writing,[9] have been commonly used. However, not all of the above methods are suitable for fabricating the diamond structure. For example, the self-assembly method is limited to the face-centered-cubic (FCC) or close-packed structures.[3] Recently, micro-manipulation has been used to fabricate the diamond structure.[4] However, the sample size is limited to a few unit cells and is very time consuming. The layer by layer and the multi-photon direct laser writing techniques had been used to fabricate the woodpile structure with bandgaps in the infrared range.[5,9] However, they are limited to a few layers due to the demanding precision and procedures.[5,9] The holographic lithography, a method combining the techniques of multiple-beam interference and the photolithography to record the interference pattern in photoresist, provides some advantages over the above methods. It requires only simple experimental setups and various structures (e.g. quasi-periodic [7] and chiral structures [8]) are feasible by different beam orientations and polarizations. This method has thus attracted much interest since the realization of the FCC structure using a 4-beam interference.[6] Furthermore, a double-exposure had also been used to fabricate the woodpile structure in the infrared range.[10] Recently, several groups have suggested that some modifications of the 4-beam configuration for the FCC structure could be used to fabricate the diamond structure.[11-14]

However, these configurations require either impractical beam arrangements with the interfering beams counter-propagating from both half spaces or elliptical polarizations that are hard to implement experimentally.[11-15] A recent attempt to fabricate the diamond structure using a (3+1)-beam configuration (3 linear polarized side beams and one circular polarized) is however debatable.[16] In this report, we propose a 5-beam configuration to fabricate the woodpile and diamond structures. The configuration is basically the "umbrella" arrangement with 4 side beams arranged symmetrically around a central beam.[6] More importantly, all the beams are from the same half space making it easy to implement experimentally. The angle between the side beams with the central beam determines the rod shape in the woodpile structure and can be chosen, with proper beam polarizations and by exploiting the advantage of the binarization using intensity cut-off,[12] to give the diamond structure. Furthermore, the central beam can be linear or circular polarized. Our configurations can be achieved using standard holographic setups, leading to possible realization of the diamond structure in the visible range.

Figure 1(a) shows the 5-beam configuration with 4 side beams $\vec{k}_n$ arranged symmetrically and making an angle $\varphi$ with a central beam $\vec{k}_0$. The wave vectors of the beams are given by:

$$\begin{bmatrix} \vec{k}_0 = k(0,0,1) \\ \vec{k}_1 = k(-\sin\varphi, 0, \cos\varphi) \\ \vec{k}_2 = k(0, -\sin\varphi, \cos\varphi) \\ \vec{k}_3 = k(\sin\varphi, 0, \cos\varphi) \\ \vec{k}_4 = k(0, \sin\varphi, \cos\varphi) \end{bmatrix}, \qquad (1)$$

where $k = 2\pi/\lambda$ and $\lambda$ is the wavelength of the light source. We define the beam polarization as the angle $\omega_i$ between the electric field $\vec{E}_i$ and the plane of incident, denoted by unit vector $\hat{n}_i$, as illustrated only for $\vec{E}_1$ in Fig. 1(a). The polarization for the central beam is defined as the angle from the x-axis also shown in Fig. 1(a). Thus, the electric fields are

$$\begin{bmatrix} \vec{E}_0 = E_0(\cos\omega_0, \sin\omega_0, 0) \\ \vec{E}_1 = E_1(\cos\omega_1\cos\varphi, \quad \sin\omega_1, \quad \cos\omega_1\sin\varphi) \\ \vec{E}_2 = E_2(-\sin\omega_2, \quad \cos\omega_2\cos\varphi, \quad \cos\omega_2\sin\varphi) \\ \vec{E}_3 = E_3(-\cos\omega_3\cos\varphi, \quad -\sin\omega_3, \quad \cos\omega_3\sin\varphi) \\ \vec{E}_4 = E_4(\sin\omega_4, \quad -\cos\omega_4\cos\varphi, \quad \cos\omega_4\sin\varphi) \end{bmatrix}. \tag{2}$$

For simplicity, we choose $\left|\vec{E}_i\right|$=1 for i =0-4. Given Eq. (1), the intensity distribution of the 5-beam interference can be expressed as

$$I(\vec{r}) = \sum_{l,m} \vec{E}_l \cdot \vec{E}_m^* e^{-i\vec{q}_{lm}\cdot\vec{r} - i(\delta_l - \delta_m)}, \tag{3}$$

where $\vec{q}_{lm} = \vec{k}_l - \vec{k}_m$ for *l,m* = 0-4 and *δ's* are the phases of the beams, chosen to be zero for simplicity. There are four independent vectors $\vec{q}_{lm}$ and any three of them, e.g. $\vec{q}_1 = \vec{k}_0 - \vec{k}_1$, $\vec{q}_2 = \vec{k}_1 - \vec{k}_2$, and $\vec{q}_3 = \vec{k}_1 - \vec{k}_4$, could be used as the reciprocal basis vectors for 3D periodic lattices. For the FCC lattice the reciprocal lattice is body-center-cubic (BCC) so that $|\vec{q}_2| = |\vec{q}_3| = \frac{2}{\sqrt{3}}|\vec{q}_1|$. This requires $\cos\varphi = 1/3$, giving $\varphi$ = 70.53°. For linear polarized beams, Eq. 3 can be simplified as

$$I(\vec{r}) = \sum_{l=m} \vec{E}_l \cdot \vec{E}_m + 2\sum_{l<m} \vec{E}_l \cdot \vec{E}_m \cos(\vec{q}_{lm}\cdot\vec{r}). \tag{4}$$

To fabricate the woodpile structure, we need rods in both the x- and y-directions stacked alternately with one "rod-space" shift in each plane. This can be done by the interference of beams $(\vec{k}_0, \vec{k}_2, \vec{k}_4)$ and $(\vec{k}_0, \vec{k}_1, \vec{k}_3)$ for the x- and y- rods shown as intensity contour surfaces in Figs. 1(a) and (b), respectively, using polarizations $\{\omega_i\} = \{45^o, 225^o, -45^o, 45^o, -45^o\}$ chosen from symmetry considerations. Note that in general the rods are elliptical. Furthermore, in order to form the woodpile structure, the y-rods in Fig. 1(c) are shifted one "rod-space" from the x-rods along the z-axis in Fig. 1(b). This is done by adding a 180° phase to one of the side beam as indicated in the polarizations used for Figs. 1(a) and (b). Figure 1(d) shows a woodpile structure obtained from the interference of all 5 beams in Fig. 1(a) using the same parameters as in Figs. 1(b)-(c). For $\varphi = 70.53^{o}$ and $\{\omega_i\}$ used in Figs.1(b)-(d), Eq. (4) can be reduced, dropping the constant term, to

$$\begin{aligned} I(\vec{r}) &\approx 2\vec{E}_0\cdot\vec{E}_1\cos(\vec{q}_{01}\cdot\vec{r}) + 2\vec{E}_0\cdot\vec{E}_3\cos(\vec{q}_{03}\cdot\vec{r}) + 2\vec{E}_1\cdot\vec{E}_3\cos(\vec{q}_{13}\cdot\vec{r}) \\ &+ 2\vec{E}_0\cdot\vec{E}_2\cos(\vec{q}_{02}\cdot\vec{r}) + 2\vec{E}_0\cdot\vec{E}_4\cos(\vec{q}_{04}\cdot\vec{r}) + 2\vec{E}_2\cdot\vec{E}_4\cos(\vec{q}_{24}\cdot\vec{r}) \end{aligned} . \qquad 5(a)$$

Column 2 in Table I give the polarizations and the intensity distributions for Eq. 5(a). Note that after a $45^{o}$ rotation along the z-axis, $I(\vec{r})$ has nearly the symmetry of a diamond structure except for the "non-diamond" cross-terms $\vec{E}_1\cdot\vec{E}_3$ and $\vec{E}_2\cdot\vec{E}_4$.[11-12] Figures 1(e) and (f) show clearly the diamond structure in unit cell for 95% and 86% intensity cut-offs, respectively. For the 86% cut-off, the contour surfaces are interconnected, a condition needed for self-supporting in the holographic lithography fabrication. Despite the "non-diamond" terms, they look remarkably the same as those reported earlier.[11-12] This is partly because the "non-diamond" terms (Fig. 1(g)) is small compared to the diamond terms (Fig. 1(h)) so that a high enough intensity cut-off would diminish the effect of the "non-diamond" terms; and also the fact that the "non-diamond" terms have the horizontal symmetry of the diamond structure such that these terms will not work "against" the diamond structure. This is a good example to demonstrate the advantage of the holographic lithography in the fabrication of photonic crystals because it is more tolerable to non-ideal beam configurations by using intensity binarization.

In principle, the "non-diamond" cross terms in Eq. 4 can be eliminated by suitable choices of the polarizations for the 5 beams. For $\varphi = 70.53^{o}$, we find one solution by imposing $\omega_0 = 45^{o}$, $\omega_1 = \omega_3 + \pi$, and $\omega_2 = \omega_4 = \pi - \omega_1$, giving $\omega_3 = \arctan(\sqrt{7/9}) = 41.41^{o}$, $\omega_1 = 221.41^{o}$, and $\omega_2 = \omega_4 = -41.41^{o}$. With these polarizations, Eq. 4 can be reduced to

$$\begin{aligned} I(\vec{r}) &\approx 2\vec{E}_0\cdot\vec{E}_1\cos(\vec{q}_{01}\cdot\vec{r}) + 2\vec{E}_0\cdot\vec{E}_3\cos(\vec{q}_{03}\cdot\vec{r}) \\ &+ 2\vec{E}_0\cdot\vec{E}_2\cos(\vec{q}_{02}\cdot\vec{r}) + 2\vec{E}_0\cdot\vec{E}_4\cos(\vec{q}_{04}\cdot\vec{r}) \end{aligned} \qquad 5(b)$$

Column 3 of Table I give the polarizations and the intensity distributions for Eq. 5(b). Now, after a 45° rotation along the z-axis, $I(\vec{r})$ has exactly the symmetry of the diamond structure.[11-12] Figures 2(a)-(d) show the diamond structure for the above configuration. Note that Figs. 2(a)-(c) look practically the same as those in Figs. 1(d)-(f), respectively. Figure 2(d) shows the "just-connected" contour surfaces in the [111] direction for many unit cells. As stated before, the interference depends very much on the relatively phases of the beams. For 4 beams interference, changes of phases only lead to translational shifts of the pattern. However for 5 beams, the phase of one of the beams has to be fixed w.r.t the others. Figures 2(e) and (f) show the interference of the 5 beams using polarizations $\{\omega_i\} = \{45^o, 41.41^o, -41.41^o, 41.41^o, -41.41^o\}$, with 95% and 60% intensity cut-offs. Note that the diamond symmetry is reduced to that of FCC. Furthermore the intensity contours do not form connected surfaces (shown as an example of separate "sheets" in Fig. 2(f) even for a 60% cut-off) until at a very low intensity cut-off. This phase dependence could lead to complications in the fabrication of the diamond structure. To demonstrate that any 3 of the 4 side beams together with the central beam $\vec{k}_0$ in Eq. 4 can be used for FCC lattice, Fig. 2(g) shows a FCC structure using the same parameters as in Figs. 2(a)-(d) but without beam $\vec{k}_1$. Other combinations produce similar FCC structures. The diamond structure can also be obtained by a double-exposure, i.e. adding the interference patterns of the x-rods and y-rods as in Figs. 1(a)-(b). In doing so some of the cross terms are automatically eliminated. Figure 2(h) shows a diamond structure using a double-exposure of beams $(\vec{k}_0, \vec{k}_2, \vec{k}_4)$ and $(\vec{k}_0, \vec{k}_1, \vec{k}_3)$ using the polarization as in Figs. 2(a)-(d). Practically, there is no difference between Figs. 2(c) and (h). The double-exposure technique could be a practical solution for non-ideal beam configurations.

The double-exposure effect can also be simulated using a (4+1)-beam configuration with a circular polarized central beam $\vec{k}_0$. In this case the polarization of the central beam $\vec{k}_0$ can be written as $\vec{E}_0 = \frac{E_0}{\sqrt{2}}(1, \quad i, \quad 0) = \vec{E}_{0x} + i\vec{E}_{0y}$ for simplicity and by choosing $\{\omega_i\} = \{-, 90^o, 90^o, 90^o, 90^o\}$ many of the cross terms in Eq. 3 are zero, giving

$$\begin{aligned} &I(\vec{r}) \approx 2\vec{E}_{0y} \cdot \vec{E}_1 \cos(\vec{q}_{01} \cdot \vec{r} + \pi/2) + 2\vec{E}_{0y} \cdot \vec{E}_3 \cos(\vec{q}_{03} \cdot \vec{r} + \pi/2) + 2\vec{E}_1 \cdot \vec{E}_3 \cos(\vec{q}_{13} \cdot \vec{r}) \\ &+ 2\vec{E}_{0x} \cdot \vec{E}_2 \cos(\vec{q}_{02} \cdot \vec{r}) + 2\vec{E}_{0x} \cdot \vec{E}_4 \cos(\vec{q}_{04} \cdot \vec{r}) + 2\vec{E}_2 \cdot \vec{E}_4 \cos(\vec{q}_{24} \cdot \vec{r}) \end{aligned} . \quad 5(c)$$

Column 4 of Table I give the polarizations and the intensity distributions for Eq. 5(c), resembling Eq. 5(a) for the linear polarized case. Figures 3(a)-(b) show the x-and y-rods obtained by the $(\vec{k}_0, \vec{k}_2, \vec{k}_4)$ and $(\vec{k}_0, \vec{k}_1, \vec{k}_3)$ beams. Note that the rods are more elliptical than those in Figs. 1(a) and (b). Figure 3(c) shows the structure obtained by the (4+1)-beam configuration with the above parameters. It looks similar to those in Figs. 1(d) and 2(a). Furthermore, the expanded views, Figs. 3(d)-(e), also are similar (except the contour surfaces are more elliptical) to those obtained for the linearly polarization cases, Figs. 1(e)-(f) and 2(b) and (c), despite the fact that the 'non-diamond" cross terms are now larger. The effect of the 'non-diamond' terms can be reduced by using larger amplitudes for the central beam.

It was pointed out recently that the use a circular polarized central beam, i.e. $\vec{E}_0 = \vec{E}_{0R} + i\vec{E}_{0I}$ with $|\vec{E}_{0R}| = |\vec{E}_{0I}|$ and $\vec{E}_{0R} \cdot \vec{E}_{0I} = 0$, is equivalent to a double-exposure in which the two patterns could have a 90$^o$ phase shift w.r.t. each other.[16] In particular, using a (3+1)-beam configuration (three linearly polarized side beams and one circular polarized central beam), the diamond structure could be formed.[16] However, in general delaying $\vec{E}_{0I}$ by 90$^o$ phase w.r.t. $\vec{E}_{0R}$ is not equivalent to shifting the pattern formed by the $\vec{E}_{0R}$ with the other side beams by the same amount. Even if such shifting effect could be achieved by appropriate choices for the beam amplitudes and polarizations, there is no guarantee the resulting pattern will have the diamond structure. As an example, Figure 3(f) shows a FCC structure using four beams $(\vec{k}_0, \vec{k}_2, \vec{k}_3, \vec{k}_4)$ with parameters as in Figs. 3(a)-(e). Note that in this case the contour surfaces are much distorted as compared to those in Figs. 1(e), 2(b) and 3(d). Similar results are obtained using other combinations of side beams with the circular polarized central beam. To further verify this subtle difference, Figs. 3(g) and (h) show contour surfaces in unit cell, showing clearly the FCC structure with long ellipsoidal motifs, at 83% disconnected and 49% connected intensity cut-offs, respectively, for a (3+1)-beam configuration using parameters as given in Ref. [16]. Despite this outcome, the idea of using circular central beam can be exploited for other applications like the fabrication of spirals as reported recently.[8]

To conclude, we show the woodpile and diamond structures can be obtained by using a 5-beam holographic interference. Either linear or circular polarized central beam can be used as compared to the elliptical polarizations reported earlier. Furthermore, all the beams of our configurations are on the same half space that can be easily achieved experimentally. However, since 5 coherent beams are involved, the phases of the beams are crucial and some control of the phases is needed. We have implemented the 5-beam interference system and results for the woodpile and diamond structures will be reported separately. [17]

**Acknowledgment**

Support from Hong Kong RGC grants CA02/03.SC01, HKUST603303, and HKUST603405 is gratefully acknowledged. The author likes to thank colleagues: Jeffrey C. W. Lee, Y. K. Pang and C. T. Chan for helpful discussions.

**Table captions**

The intensity $I(\vec{r})$ distribution for different beam configurations. Results in the last row are obtained by rotating of the pattern in the row above by 45$^{o}$ along the z-axis.

**Figure captions**

1) (a) 5-beam configuration for the woodpile structure. (b) and (c) Contour surfaces with a 83% intensity cut-off for x- and y-direction rods by interference of $(\vec{k}_0, \vec{k}_2, \vec{k}_4)$ and $(\vec{k}_0, \vec{k}_1, \vec{k}_3)$ beams using φ = 70.53$^{o}$, $\{\omega_i\} = \{45^o, 225^o, -45^o, 45^o, -45^o\}$, respectively. The x-and y- directions correspond to the $[110]$ and $[\bar{1}10]$ of the cubic lattice, respectively. (d) Contour surfaces with a 60% cut-off ($I_{max}$ = 9.21) for the 5 beams $(\vec{k}_0, \vec{k}_1, \vec{k}_2, \vec{k}_3, \vec{k}_4)$ interference using parameters as in (b) and (c). (e) and (f) Expanded views of the diamond structure in unit cell of (d) with 95% and 86% (just-connected) cut-offs, respectively. (g) Contour surfaces (cut-off intensity = 5.3, $I_{max}$ =5.4) with only the $\vec{E}_1 \cdot \vec{E}_3$ and $\vec{E}_2 \cdot \vec{E}_4$ terms. (h) Contour surfaces (cut-off intensity =8.3, $I_{max}$=8.7) of Eq. 5(a) without the terms in (g).

2) (a) Contour surfaces with 60% intensity cut-off for the 5-beam $(\vec{k}_0, \vec{k}_1, \vec{k}_2, \vec{k}_3, \vec{k}_4)$ interference using φ = 70.53° and $\{\omega_i\} = \{45^o, 221.41^o, -41.41^o, 41.41^o, -41.41^o\}$; (b) and (c) Expanded views of the diamond structure in (a) with 95% and 87% cut-offs, respectively. (d) Contour surfaces of (c) along the [111] direction of the diamond structure. (e) and (f) Contour surfaces with a 95% and 60% cut-offs for the 5 beams interference, respectively, using φ = 70.53° and $\{\omega_i\} = \{45^o, 41.41^o, -41.41^o, 41.41^o, -41.41^o\}$. (g) Contour surface with a 95% cut-off for 4-beam $(\vec{k}_0, \vec{k}_2, \vec{k}_3, \vec{k}_4)$ interference using parameters in (a). (h) Contour surfaces with 87% cut-off using parameters in Fig. 1(a) with double exposures of $(\vec{k}_0, \vec{k}_1, \vec{k}_3)$ and $(\vec{k}_0, \vec{k}_2, \vec{k}_4)$.

3) (a) and (b) Contour surfaces with a 90% intensity cut-off for x- and y- rods by interference of $(\vec{k}_0, \vec{k}_2, \vec{k}_4)$ and $(\vec{k}_0, \vec{k}_1, \vec{k}_3)$ beams using φ = 70.53° and $\{\omega_i\} = \{-, 90^o, 90^o, 90^o, 90^o\}$, respectively. Here $\vec{k}_0$ is circular polarized. (c) Contour surfaces with a 54% cut-off for the (4+1)-beam interference using parameters for (a) and (b). (d) and (e) Expanded views of the diamond structure in (c) with 90% and 60% cut-offs, respectively. (f) Contour surface with a 85% cut-off of a (3+1)-beam interference using $(\vec{k}_0, \vec{k}_2, \vec{k}_3, \vec{k}_4)$ with parameters in (a) and (b). (g) – (h) Contour surfaces with 83% and 49% cut-offs, respectively, plotted with the same orientation using parameters for Fig. 1(d) of Ref. [16] using a (3+1)-beam configuration.

Fig. 1

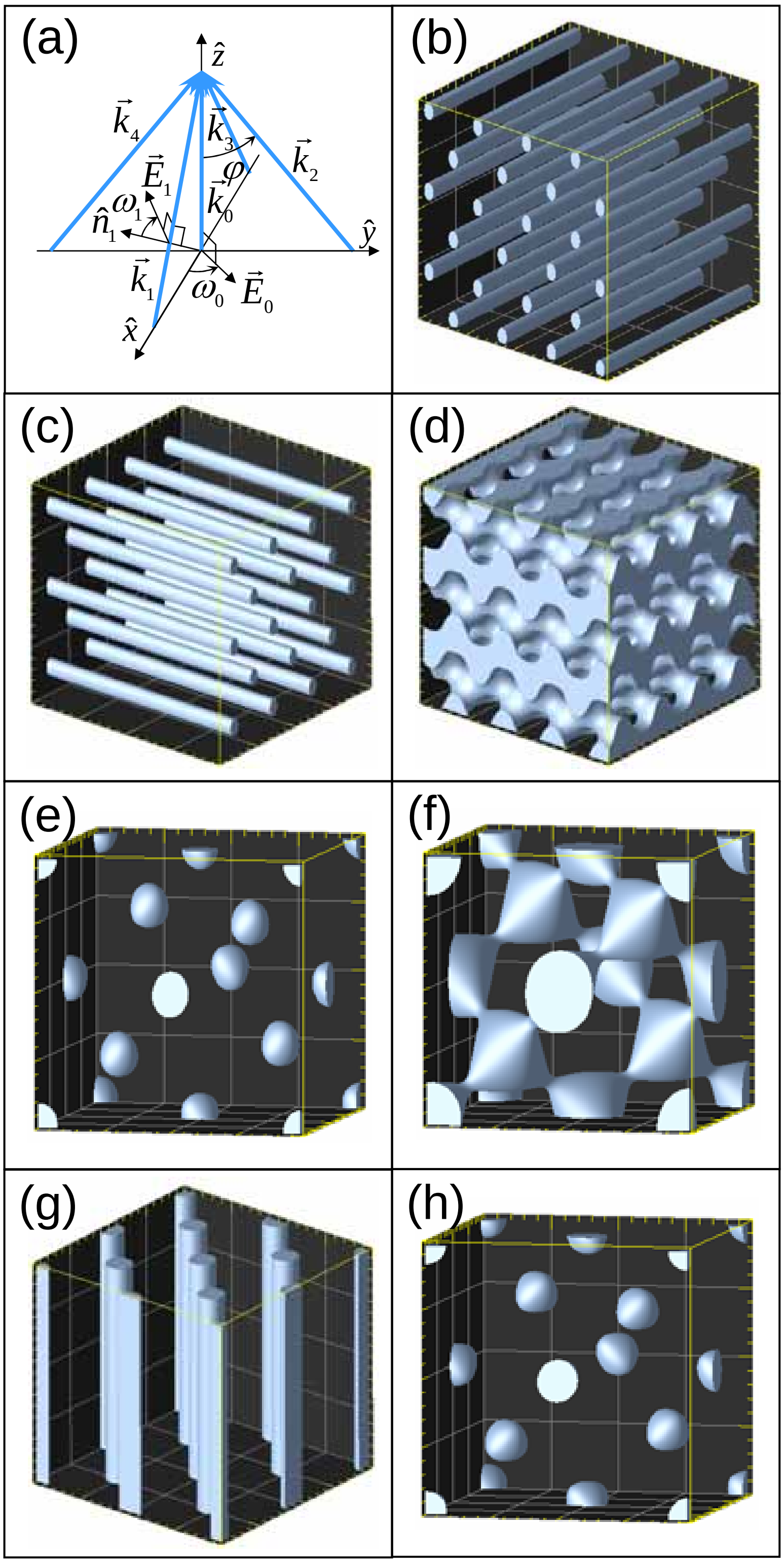

Fig. 2

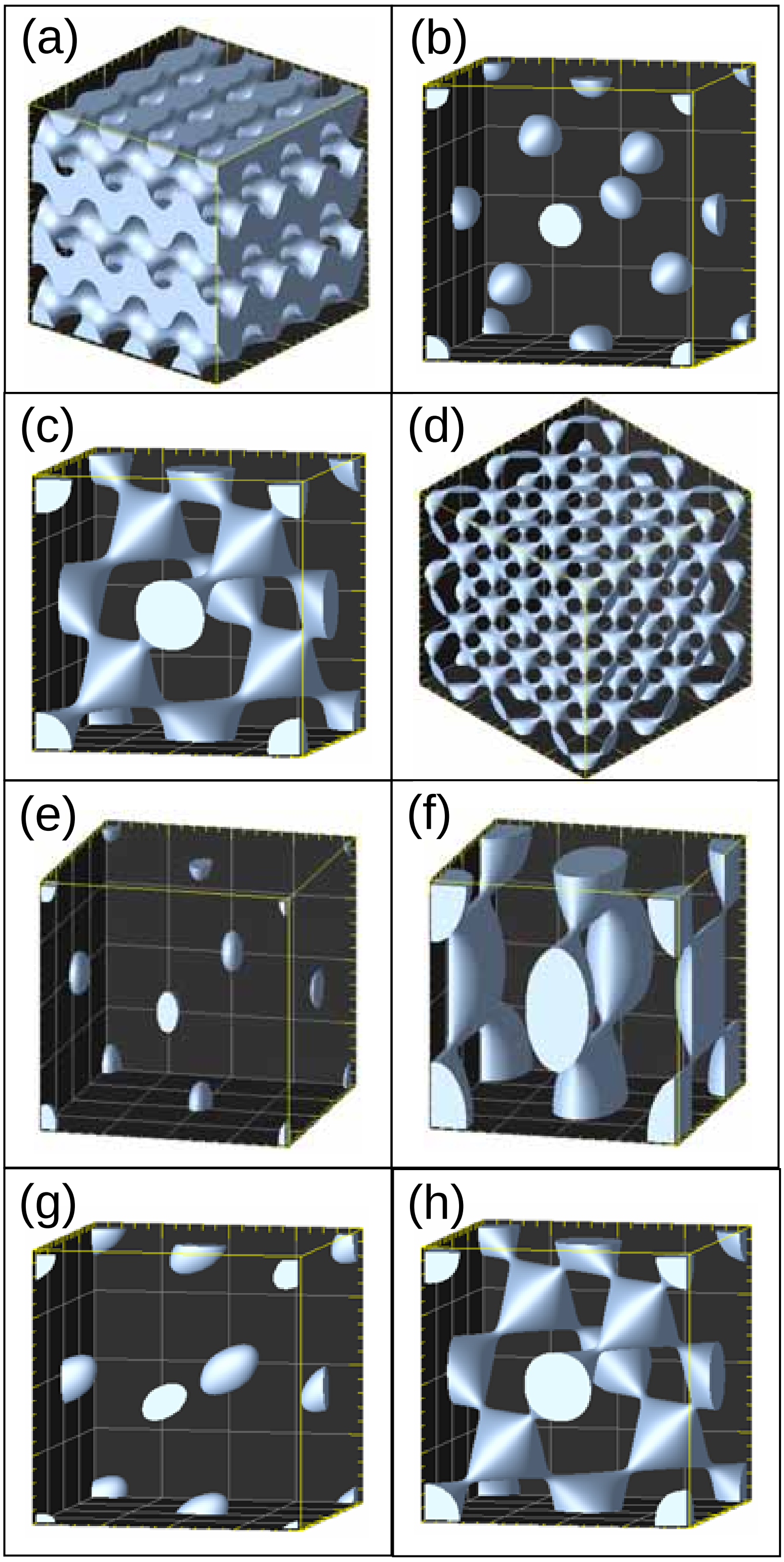

Fig. 3

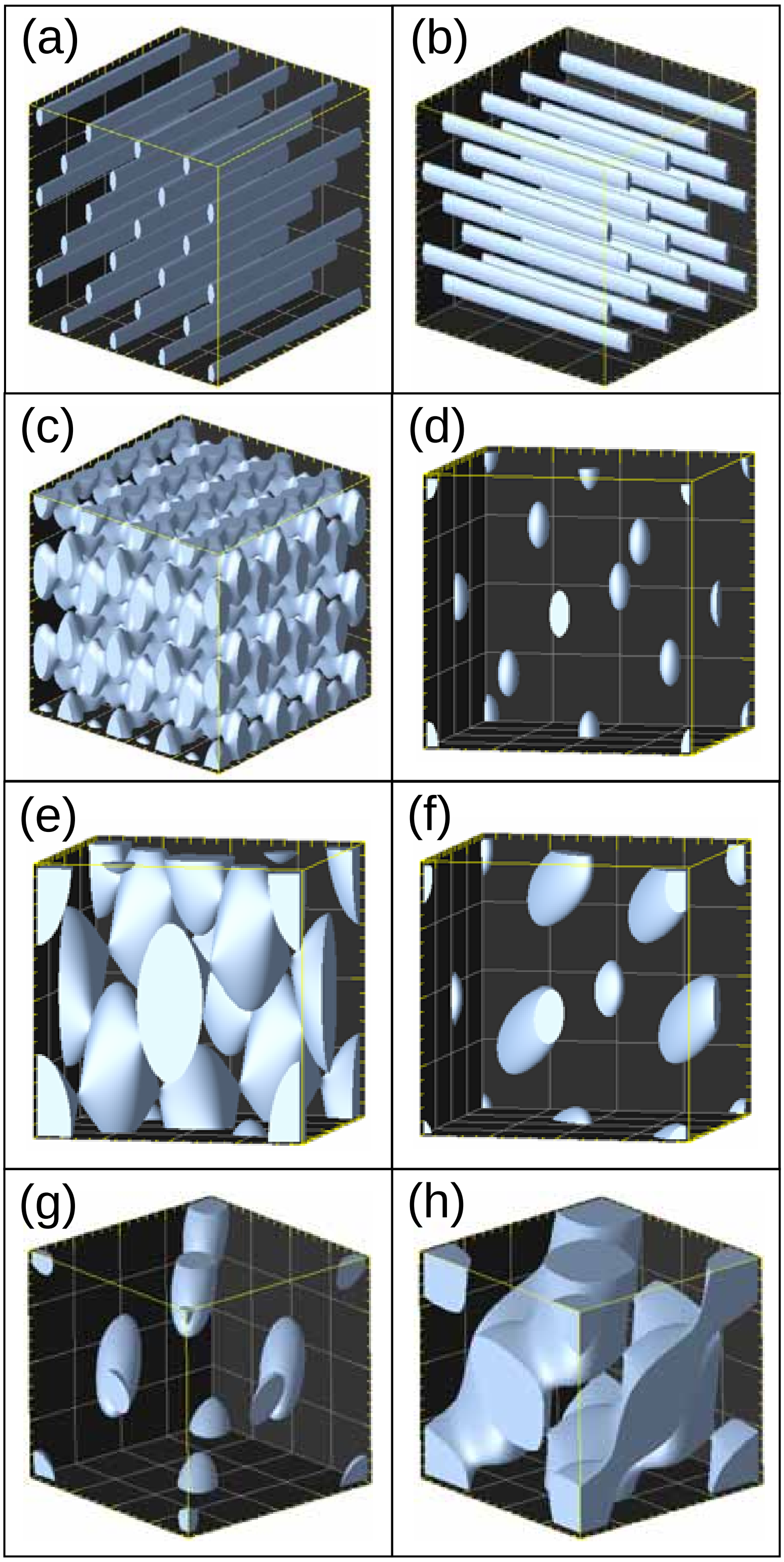

Table I

| $\vec{E}_0$ | $E_0(\cos 45^o, \sin 45^o, 0)$ | $E_0(\cos 45^o, \sin 45^o, 0)$ | $\frac{E_0}{\sqrt{2}}(1, i, 0)$ |
|---|---|---|---|
| $\{\omega_i\}$ | $\{45^o, 225^o, -45^o, 45^o, -45^o\}$ | $\{45^o, 224.41^o, -41.41^o, 41.41^o, -41.41^o\}$ | $\{-, 90^o, 90^o, 90^o, 90^o\}$ |
| $I(\vec{r})$ | $\begin{bmatrix} -\frac{4}{3}\cos(\frac{2\sqrt{2}}{3}x + \frac{2}{3}z) \\ -\frac{4}{3}\cos(\frac{-2\sqrt{2}}{3}x + \frac{2}{3}z) \\ -\frac{4}{3}\cos(\frac{2\sqrt{2}}{3}y + \frac{2}{3}z) \\ +\frac{4}{3}\cos(\frac{-2\sqrt{2}}{3}y + \frac{2}{3}z) \\ +\frac{2}{9}\cos(\frac{-4\sqrt{2}}{3}x) \\ -\frac{2}{9}\cos(\frac{-4\sqrt{2}}{3}y) \end{bmatrix}$ | $-\frac{1+\sqrt{7}}{2\sqrt{2}}\begin{bmatrix} \cos(\frac{2\sqrt{2}}{3}x + \frac{2}{3}z) \\ +\cos(\frac{-2\sqrt{2}}{3}x + \frac{2}{3}z) \\ +\cos(\frac{2\sqrt{2}}{3}y + \frac{2}{3}z) \\ -\cos(\frac{-2\sqrt{2}}{3}y + \frac{2}{3}z) \end{bmatrix}$ | $\begin{bmatrix} \sqrt{2}\cos(\frac{2\sqrt{2}}{3}x + \frac{2}{3}z + \frac{\pi}{2}) \\ -\sqrt{2}\cos(\frac{-2\sqrt{2}}{3}x + \frac{2}{3}z + \frac{\pi}{2}) \\ -\sqrt{2}\cos(\frac{2\sqrt{2}}{3}y + \frac{2}{3}z) \\ +\sqrt{2}\cos(\frac{-2\sqrt{2}}{3}y + \frac{2}{3}z) \\ -2\cos(\frac{-4\sqrt{2}}{3}x) \\ -2\cos(\frac{-4\sqrt{2}}{3}y) \end{bmatrix}$ |
| $I(R^z_{45^o}(\vec{r}))$ | $\begin{bmatrix} -\frac{4}{3}\cos(\frac{2}{3}x + \frac{2}{3}y + \frac{2}{3}z) \\ -\frac{4}{3}\cos(-\frac{2}{3}x + \frac{2}{3}y + \frac{2}{3}z) \\ -\frac{4}{3}\cos(\frac{2}{3}x + \frac{2}{3}y - \frac{2}{3}z) \\ +\frac{4}{3}\cos(\frac{2}{3}x - \frac{2}{3}y + \frac{2}{3}z) \\ +\frac{2}{9}\cos(\frac{4}{3}x + \frac{4}{3}y) \\ -\frac{2}{9}\cos(-\frac{4}{3}x + \frac{4}{3}y) \end{bmatrix}$ | $-\frac{1+\sqrt{7}}{2\sqrt{2}}\begin{bmatrix} \cos(\frac{2}{3}x + \frac{2}{3}y + \frac{2}{3}z) \\ +\cos(-\frac{2}{3}x + \frac{2}{3}y + \frac{2}{3}z) \\ +\cos(\frac{2}{3}x + \frac{2}{3}y - \frac{2}{3}z) \\ -\cos(\frac{2}{3}x - \frac{2}{3}y + \frac{2}{3}z) \end{bmatrix}$ | $\begin{bmatrix} \sqrt{2}\cos(\frac{2}{3}x + \frac{2}{3}y + \frac{2}{3}z + \frac{\pi}{2}) \\ -\sqrt{2}\cos(-\frac{2}{3}x - \frac{2}{3}y + \frac{2}{3}z + \frac{\pi}{2}) \\ -\sqrt{2}\cos(-\frac{2}{3}x + \frac{2}{3}y + \frac{2}{3}z) \\ +\sqrt{2}\cos(\frac{2}{3}x - \frac{2}{3}y + \frac{2}{3}z) \\ -2\cos(\frac{4}{3}x + \frac{4}{3}y) \\ -2\cos(\frac{4}{3}x - \frac{4}{3}y) \end{bmatrix}$ |